\documentclass[aps,prb,amsmath,amssymb,reprint,longbibliography]{revtex4-2}
\usepackage{graphicx} 
\usepackage{newtxtext} 
\usepackage{newtxmath} 
\usepackage{bm} 
\usepackage[T1]{fontenc} 
\usepackage{siunitx} 
\usepackage{xcolor}

\usepackage[colorlinks,linkcolor=blue,anchorcolor=blue,citecolor=blue,urlcolor=blue]{hyperref}

\begin{document}

\title{Magnetic Breakdown Reshapes Quantum Oscillations in Kagome Metals}

\author{Xinlong Du$^{1}$, Yuying Liu$^{1}$, Long Zhang$^{2,\ast}$ and Juntao Song$^{1,\dagger}$}
\affiliation{$^1$College of Physics and Hebei Advanced Thin Films Laboratory, Hebei Normal University, Shijiazhuang, Hebei 050024, China \\
$^2$Kavli Institute for Theoretical Sciences and School of Quantum, University of Chinese Academy of Sciences, Beijing 100190, China \\
}

\date{\today}

\begin{abstract}

Recent quantum-oscillation experiments on kagome metals have revealed markedly different phase offsets even among systems with nearly identical band structures and Fermi-surface geometries. Using a tight-binding model, we show that weak orbital hybridization can slightly modify the hybridization gaps. Small variations in these gaps can substantially alter the measured oscillation phase, despite leaving the overall electronic structure nearly unchanged. This phase shift originates from magnetic breakdown, which reconstructs cyclotron trajectories and can mask the nontrivial phase of an isolated orbit, yielding a trivial phase offset. Moreover, uniaxial strain can tune the relevant hybridization gaps and thereby weaken magnetic breakdown. This restores the nontrivial phase offset that magnetic breakdown otherwise masks, providing an experimentally accessible knob for controlling the oscillation phase. These results identify magnetic breakdown as the key mechanism controlling the phase shift and provide a plausible explanation for recent experimental phase discrepancies in kagome metals.

\end{abstract}

\maketitle
\section{\label{sec:level1} Introduction}
In condensed matter physics, the kagome lattice, with its unique band structure featuring Dirac points, van Hove singularities \cite{1}, and flat bands \cite{2,3}, provides a rich playground for exploring emergent quantum phases and exotic behaviors \cite{4,5}. Recently, quantum oscillations have proven to be a powerful experimental probe for these unconventional properties \cite{6,7}. By measuring the oscillatory components of conductivity or resistivity as a function of the inverse magnetic field ($1/B$), one can directly access the Fermi surface geometry \cite{8,9} and extract key information about the underlying quantum states, including their nontrivial topological properties \cite{10,11,12,13}. In quantum oscillations, such topological information is reflected by the Berry phase accumulated along a closed quasiparticle cyclotron orbit. Within the semiclassical quantization condition, the area enclosed by a closed cyclotron orbit satisfies \cite{14,15}
\begin{equation}
\label{eq1}
A=\frac{2\pi eB}{\hbar}(n+\gamma),
\qquad n\in\mathbb{N},
\end{equation}
where $n$ is the Landau-level index, and the constant $\gamma$ is directly related to the Berry phase $\phi_B$ accumulated along the closed cyclotron orbit, $2\pi(\gamma-1/2)=-\phi_B$. The phase factor $\gamma$ determines the positions of peaks and dips in the quantum-oscillation signal and can therefore be extracted from a Landau-fan analysis \cite{16}. Within a fixed indexing convention, a $\pi$ Berry phase is usually associated with a nontrivial phase offset, whereas a near-zero Berry phase corresponds to a trivial phase offset.

In the widely studied AV$_3$Sb$_5$ (A = K, Rb, Cs) kagome family \cite{17,18,19,20,21,22,23,24,25}, quantum oscillation measurements have played a key role in revealing the interplay between multiple electronic instabilities and topologically nontrivial band structures \cite{26,27,28,29,30,31,32}. More recently, the titanium-based analogs ATi$_3$Bi$_5$ (A = Rb, Cs) have attracted significant attention because of their distinct electronic properties \cite{33,34,35,36,37}. Compared with the extensively studied AV$_3$Sb$_5$ family \cite{38,39}, ATi$_3$Bi$_5$ has emerged as a comparatively clean platform for probing intrinsic kagome physics and quantum oscillations \cite{40,41,42}. In this isostructural family, CsTi$_3$Bi$_5$ and RbTi$_3$Bi$_5$ are expected to display similar oscillatory behavior because first-principles calculations indicate that they possess highly similar band structures and Fermi-surface geometries. However, recent quantum-oscillation measurements reveal markedly different phase offsets in these two compounds \cite{43}. This phase discrepancy raises a more general theoretical question: can a weak hybridization scale, even when it leaves the overall band structure and Fermi-surface geometry nearly unchanged, produce a large change in the phase offset of high-field quantum oscillations? In multiband kagome systems, this question is naturally connected to magnetic breakdown \cite{44,45}. Weak hybridization controls the small gaps, and these gaps determine the probability for quasiparticles to tunnel between neighboring orbits under a strong magnetic field. When such tunneling is suppressed, quasiparticles follow an isolated closed orbit. When magnetic breakdown becomes active, however, the cyclotron motion is reconstructed into composite trajectories involving multiple orbit segments. The Berry phases accumulated along these segments can add or partially cancel, so the measured oscillation phase can differ substantially from that expected for an isolated Fermi pocket.

\begin{figure}[htbp]
    \centering
    \includegraphics[width=0.48\textwidth]{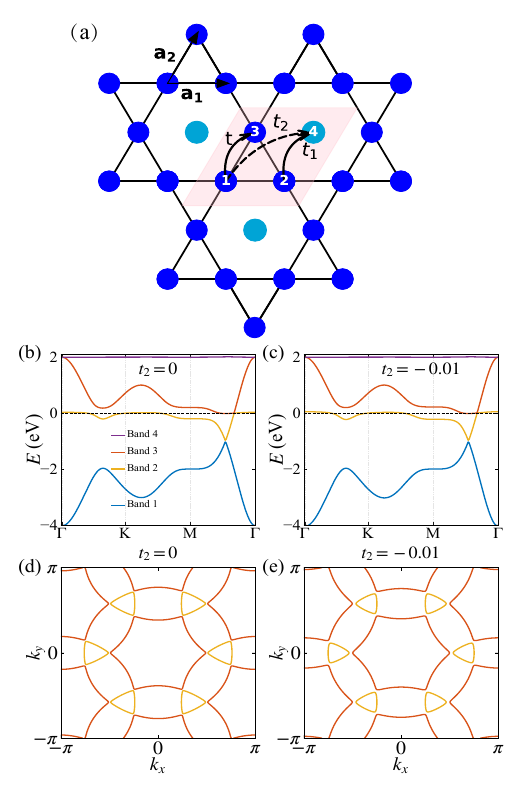}
    \caption{Lattice schematic and electronic structure of the kagome lattice. (a) Schematic of the kagome lattice with nearest-neighbor ($t$, $t_1$) and next-nearest-neighbor ($t_2$) hoppings. (b,c) Calculated band structures with spin-orbit coupling ($\lambda=0.05t$) for (b) the reference case ($t_2=0$) and (c) the weak-hybridization case ($t_2=-0.01$). (d,e) Corresponding Fermi-surface geometries at $E_F=0$ for (d) $t_2=0$ and (e) $t_2=-0.01$. A finite $t_2$ slightly enlarges the interband gap while leaving the overall Fermi-surface geometry nearly unchanged.}
    \label{fig:1}
\end{figure}

Motivated by this physical picture, we construct a tight-binding toy model, as illustrated in Fig.~\ref{fig:1}, to isolate the generic role of weak orbital hybridization in high-field quantum oscillations. Here, $t$ denotes the nearest-neighbor hopping between kagome sites, $t_1$ describes the nearest-neighbor coupling between the central site and the kagome sites, and $t_2$ is introduced as a weak next-nearest-neighbor hopping between the central and kagome sites. We further include an effective complex second-neighbor chiral hopping term to generate the gapped topological band structure relevant to the present study. In this setting, $t_2$ serves as a controlled theoretical knob for testing how weak orbital hybridization modifies avoided-crossing gaps, magnetic-breakdown connectivity, and the resulting phase offset of quantum oscillations. Our calculations show that a perturbative change in $t_2$ can substantially modify the quantum-oscillation spectra and Landau-fan intercepts by tuning magnetic-breakdown dynamics. We further demonstrate that uniaxial strain provides a tuning knob for controlling this orbit connectivity. These results reveal a general mechanism by which weak orbital hybridization can substantially modify magnetic-breakdown trajectories and thereby alter the extracted oscillation phase.

The rest of this paper is organized as follows: In Sec.~\ref{sec:level2}, we introduce the theoretical model and its Hamiltonian, together with the methodology for calculating the density of states. Sec.~\ref{sec:level3} presents the numerical results and the physical mechanism of magnetic breakdown, including the effects of uniaxial strain. Finally, a summary of our findings is provided in Sec.~\ref{sec:level4}.

\section{\label{sec:level2}Model and Method}

\begin{figure*}[htbp]
    \centering
    \includegraphics[width=0.96\textwidth]{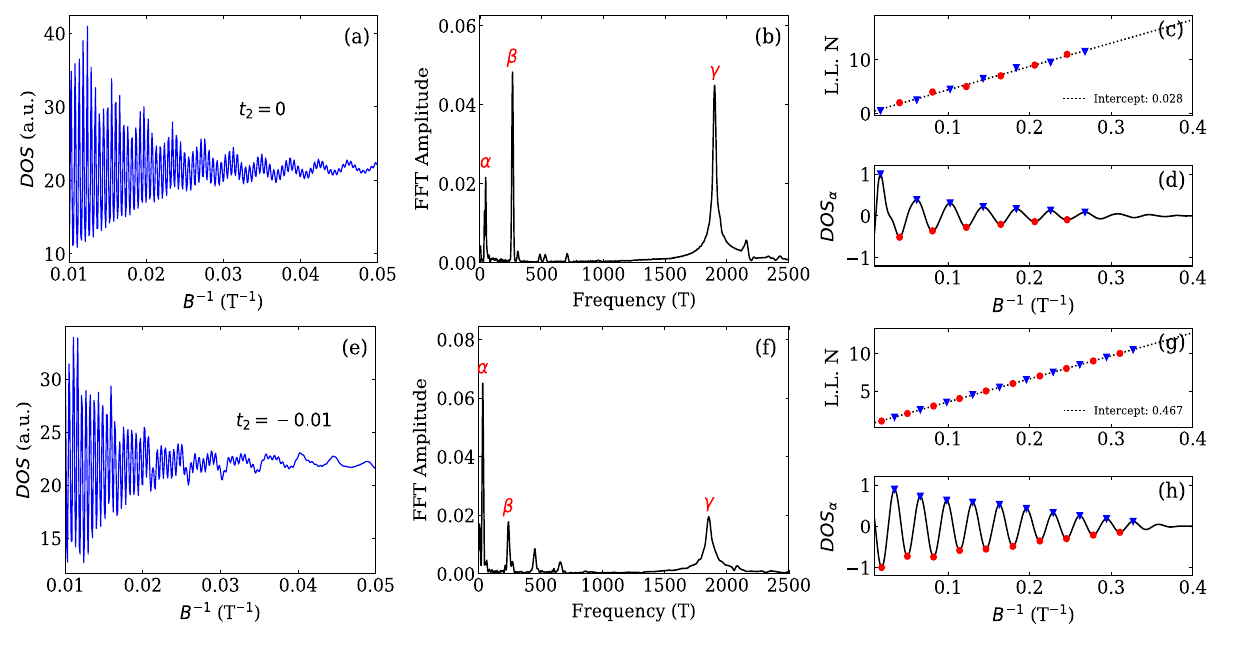}
    \caption{Quantum oscillation signatures and phase-offset extraction. (a, e) DOS oscillations at $E_F=0$ versus $1/B$ with the effective chiral hopping term ($\lambda=0.05t$) for the reference case ($t_2=0$) and the weak-hybridization case ($t_2=-0.01$), respectively. (b, f) Corresponding FFT spectra. The dominant branches are identified as $\alpha$, $\beta$, and $\gamma$, with frequencies $(46.3, 264.1, 1901.2)$ T for $t_2=0$ and $(34.3, 243.2, 1851.4)$ T for $t_2=-0.01$. (d, h) Filtered DOS isolating the $\alpha$-frequency oscillations, enabling precise identification of the physical extrema. (c, g) Landau fan diagrams constructed from the extrema of the filtered $\alpha$ oscillations. The linear extrapolations yield intercepts of 0.028 for the reference case ($t_2=0$) and 0.467 for the weak-hybridization case ($t_2=-0.01$), consistent with near-zero and $\pi$-like phase offsets, respectively.}
    \label{fig:2}
\end{figure*}

To investigate how weak orbital hybridization reshapes the phase offset of quantum oscillations, we construct a tight-binding lattice model on the kagome lattice [Fig.~\ref{fig:1}(a)] \cite{46}. The Hamiltonian is written as
\begin{equation}
\label{eq:Htot}
H = H_0 + H_{\mathrm{SOC}},
\end{equation}
with
\begin{equation}
H_0
=
t\sum_{\langle ij\rangle}a_i^\dagger a_j
+
t_1\sum_{\langle ij\rangle}a_i^\dagger b_j
+
t_2\sum_{\langle\langle ij\rangle\rangle}a_i^\dagger b_j
\end{equation}
and
\begin{equation}
H_{\mathrm{SOC}}
=
i\frac{2\lambda}{\sqrt{3}}
\sum_{\langle\langle ij\rangle\rangle}
(\mathbf d_{ij}^1\times \mathbf d_{ij}^2)\cdot\boldsymbol\sigma\,
a_i^\dagger a_j .
\end{equation}

Here, $a_i=(a_{i\uparrow}, a_{i\downarrow})^T$ and $b_i=(b_{i\uparrow},b_{i\downarrow})^T$ denote electron annihilation spinors on the kagome sites (blue) and central sites (cyan), respectively. The parameter $t$ describes nearest-neighbor hopping within the kagome network, and $t_1$ and $t_2$ describe nearest- and next-nearest-neighbor hopping between the central and kagome sites, respectively. The spin-orbit term $H_{\mathrm{SOC}}$ acts on second-neighbor hoppings between kagome sites. For a hopping path from site $j$ to site $i$, $\mathbf d_{ij}^{1}$ and $\mathbf d_{ij}^{2}$ denote the two nearest-neighbor bond vectors along the corresponding two-step path. Their cross product defines the local chirality of this path, with opposite signs for clockwise and counterclockwise hopping processes. The Pauli matrices $\boldsymbol\sigma$ act in this spin space; since the bond vectors lie in the kagome plane, the cross product points along the $z$ direction and couples to $\sigma_z$. Together with the prefactor $i$, this chirality factor produces a spin-dependent complex second-neighbor hopping, analogous to the Haldane-type hopping pattern \cite{47}. This term opens a topologically nontrivial gap in the relevant band sector and produces the Berry curvature needed for the phase analysis \cite{48}. Further details are given in Appendix~\ref{appendixB}.

This model is designed to examine the dynamical consequences of weak orbital hybridization while keeping the overall band structure and Fermi-surface geometry nearly unchanged. As shown in Figs.~\ref{fig:1}(b)--1(e), we compare the reference case $t_2=0$ with the weak-hybridization case $t_2=-0.01$. Throughout this work, we set $t=1$ as the energy unit, fix $t_1=-0.1$, and choose $\lambda=0.05t$. The chosen parameter set satisfies $|t_2| \ll |t_1| \ll |t|$, so that $t_2$ acts as a weak perturbation to the electronic structure. This choice allows us to tune the hybridization gaps while leaving the overall band structure and Fermi-surface geometry nearly unchanged. In this way, the model provides a controlled setting for analyzing how weak hybridization modifies magnetic-breakdown connectivity in high-field quantum oscillations.

To calculate quantum oscillations, we introduce a perpendicular magnetic field $B$ through the Peierls substitution,
\begin{equation}
t_{ij}\rightarrow t_{ij}e^{i\theta_{ij}},\qquad
\theta_{ij}=\frac{e}{\hbar}\int_i^j \mathbf A\cdot d\mathbf l
=\frac{2\pi}{\Phi_0}\int_i^j \mathbf A\cdot d\mathbf l,
\end{equation}
where $\mathbf A=(-yB,0,0)$ and $\Phi_0=h/e$ is the flux quantum \cite{49,50,51}. Numerical calculations are performed on a two-dimensional lattice with linear dimensions $N_x=N_y=200a$, and all energies are measured in units of $t$.

For each magnetic field, we calculate the density of states (DOS) at the Fermi level from the retarded Green's function \cite{53},
\begin{equation}
DOS(E)
=
-\frac{1}{\pi}\operatorname{Im}\operatorname{Tr}G^r(E),
\end{equation}
where $G^r(E)=(E-H+i\eta)^{-1}$ and $\eta=0.001t$ is a small broadening parameter. The oscillatory part of $DOS(E_F)$ as a function of $1/B$ is then used to extract the quantum-oscillation spectrum and the corresponding Landau-fan phase offset. In the numerical implementation, the lattice is divided into slices, so that the Hamiltonian has a block-tridiagonal form. The required diagonal Green's-function blocks are obtained efficiently using the recursive Green's function method, as detailed in Appendix~\ref{appendixC}.

\begin{figure*}[htbp]
    \centering
    \includegraphics[width=0.98\textwidth]{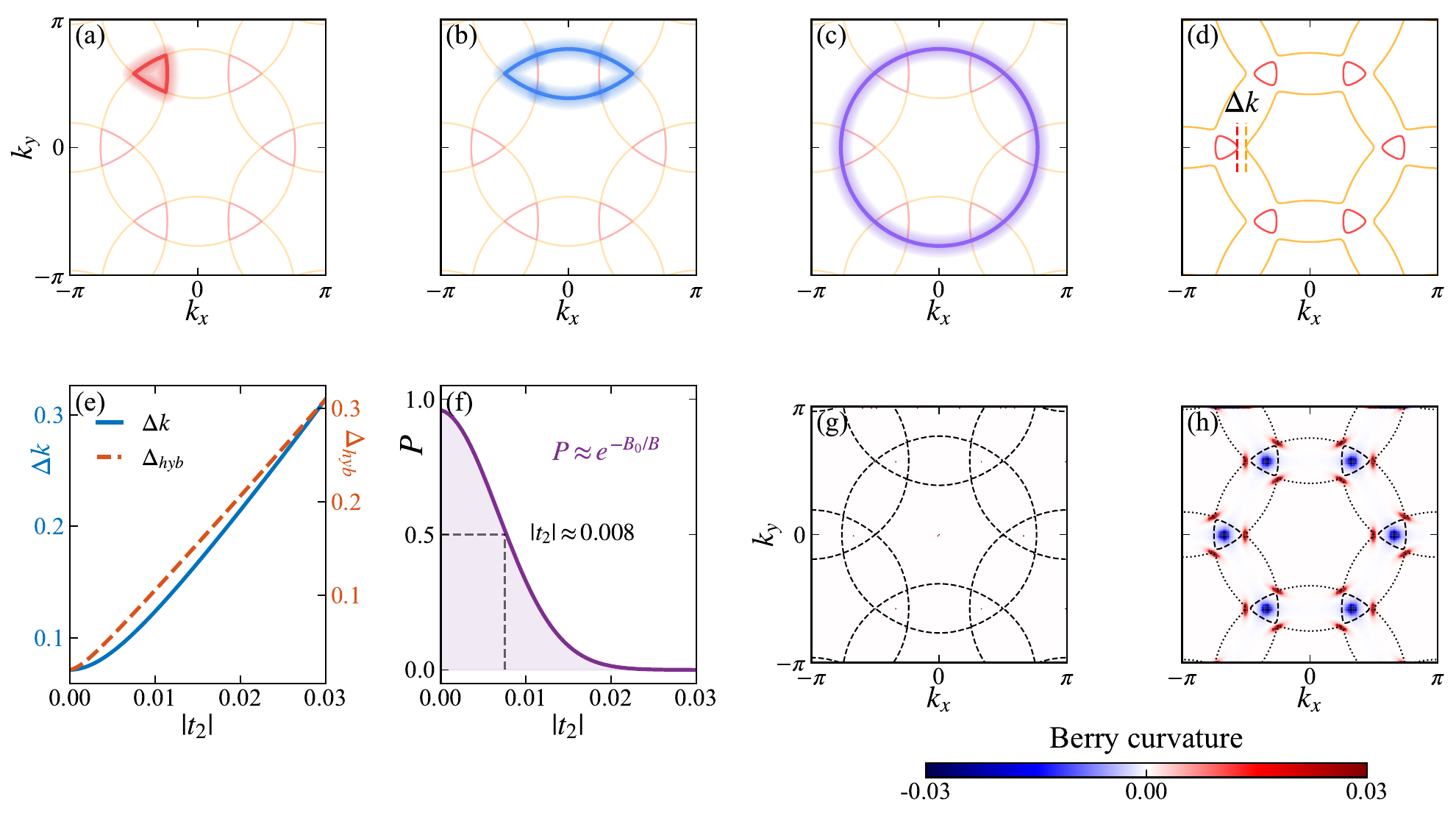}
    \caption{Mechanism of magnetic breakdown (MB) quenching and geometric-phase visibility. (a) Isolated $\alpha$ orbit without MB. (b,c) Magnetic-breakdown trajectories corresponding to the partial-breakdown $\beta$ orbit and the full MB $\gamma$ orbit. (d) Definition of the momentum separation $\Delta k$ between neighboring closed orbits. (e) Evolution of $\Delta k$ and the hybridization gap $\Delta_{\mathrm{hyb}}$ with $|t_2|$. (f) Magnetic-breakdown probability $P$ at $B=35$~T, showing a rapid crossover near $|t_2| \approx 0.008$. (g,h) Berry-curvature distributions associated with the orbit reconstruction. (g) Summed Berry curvature, $\Omega_2+\Omega_3$, of the two adjacent bands in the strong-breakdown case ($t_2=0$), showing near-cancellation of the dominant local geometric contributions. (h) Band-3-resolved Berry curvature at $t_2=-0.01$, showing that the isolated triangular pocket encloses a net Berry flux, consistent with the $\pi$-like phase offset extracted for the $\alpha$ branch. Dashed contours mark the corresponding $E_F=0$ orbits.}
    \label{fig:3}
\end{figure*}
\section{\label{sec:level3}Numerical Results}
\subsection{Quantum Oscillation Signatures}
In both cases, spin-orbit coupling keeps the relevant band topology nontrivial (Fig.~\ref{fig:S1}). Their similar band structures and Fermi-surface geometries would therefore suggest comparable phase offsets. To examine this expectation, we compute the Landau-quantized DOS and analyze the resulting quantum oscillations [Figs.~\ref{fig:2}(a) and \ref{fig:2}(e)]. The oscillation frequencies provide a direct probe of the underlying orbit areas, while the Landau-fan intercepts reveal the corresponding phase offsets.

The fast Fourier transform (FFT) spectra reveal that the three dominant branches, denoted by $\alpha$, $\beta$, and $\gamma$, persist in both cases ($t_2 = 0$ and $t_2 = -0.01$). For $t_2=0$, the corresponding frequencies are $46.3$~T, $264.1$~T, and $1901.2$~T, whereas for $t_2=-0.01$ they shift to $34.3$~T, $243.2$~T, and $1851.4$~T, respectively [Figs.~\ref{fig:2}(b) and \ref{fig:2}(f)]. The persistence of these branches indicates that the overall oscillation spectrum remains organized by a similar orbit network in the two cases, although their frequencies shift with $t_2$. More importantly, the relative spectral weights are strongly redistributed: in the case of $t_2=0$, the intermediate $\beta$ branch dominates the spectrum, whereas in the weak-hybridization case ($t_2=-0.01$), the dominant weight is transferred to the fundamental $\alpha$ branch. Taken together, these results show that weak hybridization has a pronounced effect on the observable spectral-weight distribution.

We next analyze the Landau-fan diagrams to extract the phase offsets. To obtain the fan diagrams, we first isolate the fundamental $\alpha$ oscillations through frequency filtering [Figs.~\ref{fig:2}(d) and \ref{fig:2}(h)]. Using the indexing convention associated with Eq.~(\ref{eq1}), we assign integer indices $n$ to successive maxima of the filtered $\alpha$ oscillations. For each maximum, we record its magnetic-field position $B_n$ and plot $n$ as a function of $1/B_n$. The data are then fitted linearly as $n=F(1/B_n)+n_0$, where the slope gives the oscillation frequency $F$ and the intercept $n_0$ gives the Landau-fan offset. Within this indexing convention, $n_0\simeq 0$ and $n_0\simeq 0.5$ correspond to phase offsets close to 0 and $\pi$, respectively.

The resulting Landau-fan diagrams reveal a clear discrepancy between the extracted phase offset and the phase expected from the nontrivial band topology [Figs.~\ref{fig:2}(c) and \ref{fig:2}(g)]. In the $t_2=0$ case, the linear fit yields $n_0=0.028$, consistent with a trivial phase offset. In contrast, introducing weak hybridization ($t_2=-0.01$) shifts the intercept to $n_0=0.467$, corresponding to a $\pi$-like phase offset. Thus, the weak-hybridization case ($t_2=-0.01$) exhibits the expected nontrivial phase offset. However, the $t_2=0$ case does not, despite having the same nontrivial band topology.

For completeness, we also analyzed the Landau-fan diagrams for the higher-frequency $\beta$ and $\gamma$ branches (see Appendix~\ref{appendixD}). In contrast to the $\alpha$ branch, their fitted intercepts remain close to zero in both hybridization cases, consistent with near-trivial phase offsets. The central question is therefore why the nontrivial phase offset becomes visible only for the $\alpha$ oscillations in the case of $t_2=-0.01$. The $\beta$ and $\gamma$ branches remain near-trivial despite the same topological character. Because the introduction of $t_2$ leaves the overall Fermi-surface geometry nearly unchanged, this behavior cannot be explained by the conventional Onsager relation, which depends only on static extremal orbits. These results suggest that the phase shift likely originates from the magnetic-field-induced reconstruction of semiclassical trajectories through interband magnetic breakdown.

\subsection{Magnetic Breakdown Mechanism}
\begin{figure*}[htbp]
    \centering
    \includegraphics[width=0.80\textwidth]{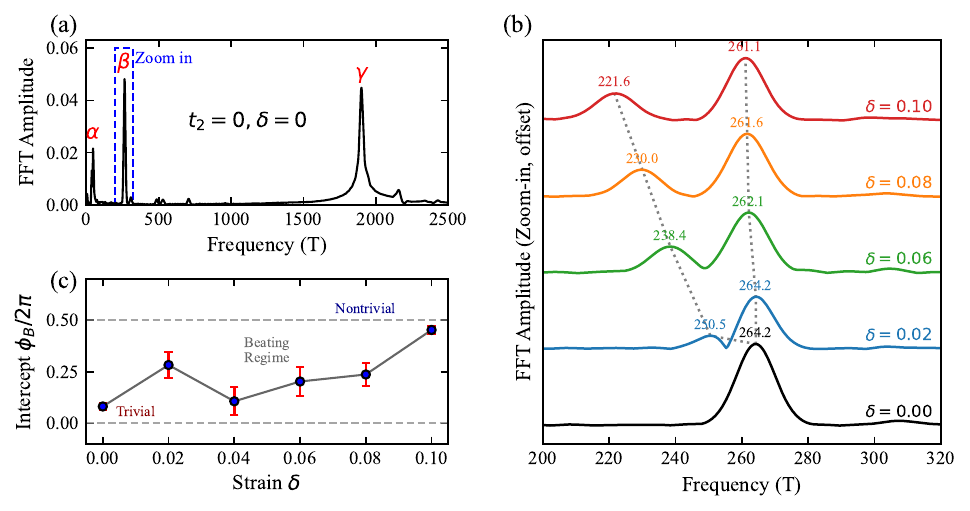}
    \caption{Strain-dependent evolution of the oscillation spectrum and phase offset in the strong-breakdown reference regime ($t_2=0$). (a) Global FFT spectrum of the unstrained system ($\delta=0$). (b) Representative FFT evolution of the $\beta$ branch under four uniaxial strain levels, $\delta=0.02$, $0.06$, $0.08$, and $0.10$, showing its splitting into the inequivalent branches $\beta_1$ and $\beta_2$. (c) Extracted Landau-fan intercepts as a function of strain $\delta$, revealing a crossover from a near-zero-intercept strong-breakdown regime, through an intermediate beating regime, toward the $\pi$-like isolated-orbit limit at larger strain.}
    \label{fig:4}
\end{figure*}

To clarify the role of magnetic breakdown in the phase-offset discrepancy, we first compare the frequencies expected from the enclosed Fermi-surface areas with those in the FFT spectra. From the constant-energy contours of the momentum-space Hamiltonian at $E_F$, we find only two fundamental closed Fermi-surface pockets [Fig.~\ref{fig:3}(a)]: a small triangular pocket highlighted in red and a larger central hexagonal pocket traced by the yellow contour. Through the Onsager relation, their enclosed areas correspond to oscillation frequencies of approximately $46$~T and $610$~T, respectively. However, the FFT spectra [Figs.~\ref{fig:2}(b,f)] show a different frequency structure. The low-frequency $\alpha$ branch associated with the triangular pocket remains visible, whereas the $610$~T signature expected from the central hexagonal pocket is absent. Instead, two additional branches, $\beta$ and $\gamma$, emerge in the spectra. The $\beta$ and $\gamma$ frequencies can be mapped onto the composite semiclassical trajectories shown in Figs.~\ref{fig:3}(b) and \ref{fig:3}(c), with deviations below $3\%$. Specifically, the $\beta$ branch corresponds to the spindle-shaped orbit [Fig.~\ref{fig:3}(b)], and the $\gamma$ branch maps to the large enveloping composite orbit [Fig.~\ref{fig:3}(c)]. Since the orbit areas inferred from the $\beta$ and $\gamma$ frequencies do not correspond to independent Fermi-surface pockets, this indicates that the electronic trajectories are reconstructed by magnetic breakdown under a strong magnetic field.

To test this magnetic-breakdown interpretation, we numerically evaluate the dependence of the magnetic-breakdown probability on the tuning parameter $t_2$. Magnetic breakdown is highly sensitive to the hybridization gap, and the tunneling probability is approximately given by \cite{52}
\begin{equation}
P \approx e^{-B_0/B},
\label{eq8}
\end{equation}
where $B_0$ is the characteristic breakdown field. For fixed local band velocities near the avoided crossing, $B_0$ increases approximately with $\Delta_{\mathrm{hyb}}^2$. Figures~\ref{fig:3}(d) and \ref{fig:3}(e) show that increasing $|t_2|$ not only enlarges the momentum-space separation $\Delta k$ between neighboring closed orbits, but also increases the corresponding hybridization gap $\Delta_{\mathrm{hyb}}$. According to Eq.~\eqref{eq8}, increasing $|t_2|$ is expected to suppress the magnetic-breakdown probability exponentially. This trend is confirmed in Fig.~\ref{fig:3}(f), where a rapid crossover from strong to weak magnetic breakdown occurs near $|t_2| \approx 0.008$. Thus, the weak finite value $t_2=-0.01$ is sufficient to strongly suppress magnetic breakdown.

This suppression provides a direct test of the origin of the $\beta$ and $\gamma$ branches. If these branches are produced by magnetic breakdown, their spectral weight should be largest in the strong-breakdown regime and should be strongly reduced once the tunneling probability is suppressed. This is precisely what is seen in the numerical spectra. In the $t_2=0$ case, the $\beta$ and $\gamma$ branches carry the dominant spectral weight [Fig.~\ref{fig:2}(b)]. For $t_2=-0.01$, their amplitudes are strongly attenuated, while the fundamental $\alpha$ branch becomes dominant [Fig.~\ref{fig:2}(f)]. This spectral-weight evolution provides direct evidence that the $\beta$ and $\gamma$ branches are magnetic-breakdown orbits generated under a strong magnetic field, rather than independent Fermi-surface pockets.

The above analysis establishes two contrasting magnetic-breakdown regimes. At $t_2=0$, strong magnetic breakdown reconstructs the semiclassical trajectories, whereas at $t_2=-0.01$ magnetic breakdown is strongly suppressed. This contrast suggests that the different phase offsets originate from the different magnetic-breakdown connectivity in the two cases. To clarify this point, we analyze the Berry-curvature distribution in momentum space.

For an isolated closed orbit on band $m$, the accumulated geometric phase $\phi_B$ can be written as the Berry flux through the enclosed momentum-space area $S$ \cite{14},
\begin{equation}
\phi_B = \iint_S \left[ \nabla_{\mathbf{k}} \times \mathbf{A}_m(\mathbf{k}) \right] \cdot d\mathbf{S}
= \iint_S \mathbf{\Omega}_m(\mathbf{k}) \cdot d\mathbf{S},
\end{equation}
where $m$ denotes the band index, $\mathbf{A}_m(\mathbf{k})$ is the Berry connection, and $\mathbf{\Omega}_m(\mathbf{k})$ is the Berry curvature. The local contributions to this surface integral are shown in Figs.~\ref{fig:3}(g) and \ref{fig:3}(h). The Berry curvature is strongly concentrated near the avoided crossings, forming localized hotspots. For bands 2 and 3 [Fig.~\ref{fig:1}(b)], these hotspots carry opposite signs at nearly the same momentum-space locations.

This opposite-sign Berry curvature provides the physical origin of the phase dichotomy. In the strong-breakdown regime ($t_2=0$), the small hybridization gap allows tunneling between neighboring orbits. As shown in Fig.~\ref{fig:3}(g), the Berry-curvature contributions from the two adjacent bands largely cancel each other. The resulting net Berry flux is therefore strongly reduced, leading to a near-trivial phase offset in the oscillation signal.

In contrast, a weak finite hopping $t_2=-0.01$ enlarges the gaps at the avoided crossings and suppresses magnetic breakdown. Interband tunneling is strongly attenuated, confining the cyclotron motion to the orbit associated with the isolated fundamental pocket [Fig.~\ref{fig:3}(a)]. As shown in Fig.~\ref{fig:3}(h), this isolated trajectory is no longer subject to cancellation between adjacent bands. Instead, it encloses a net Berry flux from a single band, yielding the nontrivial $\pi$-like phase offset extracted from the Landau-fan analysis.

These results establish the central mechanism of this work: weak orbital hybridization does not change the band topology in the parameter regime considered here, but controls the measured oscillation phase by tuning magnetic-breakdown connectivity between neighboring orbits. The oscillation phase is therefore governed not only by the underlying electronic structure, but also by magnetic-field-induced tunneling between adjacent orbits. This mechanism provides a possible explanation for why closely related kagome metals can display markedly different phase offsets despite having highly similar band structures.

\subsection{Strain-Driven Recovery of the Nontrivial Phase Offset}

Having identified magnetic-breakdown connectivity as the key factor controlling the oscillation phase, we next ask whether the phase offset can be tuned by an independent external perturbation. To this end, we introduce uniaxial strain as a continuous tuning parameter starting from the strong-breakdown case ($t_2=0$) \cite{54}. The strain dependence is implemented through the anisotropic hopping model described in Appendix~\ref{appendixE}. In this way, magnetic-breakdown connectivity can be tuned continuously, providing an experimentally accessible route for controlling the oscillation phase offset.

The FFT spectrum in the unstrained limit is shown in Fig.~\ref{fig:4}(a). In this case, strong magnetic breakdown between neighboring orbits makes the $\beta$ branch dominant in the spectrum. As the uniaxial strain increases, the spectrum evolves substantially. For representative strain values ($\delta=0.02$, $0.06$, $0.08$, and $0.10$), the original $\beta$ branch splits into two inequivalent branches, denoted $\beta_1$ and $\beta_2$ [Fig.~\ref{fig:4}(b)]. Their strain dependences are distinct: the $\beta_1$ peak shifts continuously to lower frequency, whereas the $\beta_2$ peak remains nearly fixed near $260$~T.

The microscopic origin of this splitting can be identified from the strain-distorted Fermi surface [Fig.~\ref{fig:S4}]. The $\beta_1$ branch is associated with the upper pocket, whose enclosed area is strongly reduced under strain. Its frequency therefore shifts downward according to the Onsager relation. By contrast, the $\beta_2$ orbit is affected mainly through a shape distortion, while its enclosed area changes only weakly, leaving its frequency nearly fixed. The resulting orbit-selective splitting produces two closely spaced $\beta$-derived frequencies and accounts for the beating pattern in the strained regime.

Figure~\ref{fig:4}(c) summarizes the strain dependence of the fitted Landau-fan intercept. In the unstrained limit ($\delta=0$), strong magnetic breakdown across the composite trajectory yields a near-trivial intercept ($\approx 0.082$). At weak strain ($\delta=0.02$), the system enters an intermediate beating regime, where the fitted intercept shows nonmonotonic variations. This behavior does not indicate a topological transition. Instead, it reflects partial magnetic breakdown: interference between the split $\beta_1$ and $\beta_2$ trajectories makes the phase extraction ambiguous. At larger strain, the hybridization gaps at the avoided crossings increase, and tunneling between the orbit segments forming the $\beta_1$ branch is progressively suppressed. The cancellation among Berry-phase contributions from different segments is then weakened, allowing a net geometric phase to accumulate along the remaining stable trajectory. As a result, the fitted intercept of the $\beta_1$ branch moves toward the $\pi$ limit. This recovery occurs without changing the band topology, indicating that the near-trivial phase at $\delta=0$ arises from magnetic-breakdown-induced Berry-phase cancellation.

These results show that the quantum-oscillation phase offset can be tuned continuously in the present model. More generally, uniaxial strain provides a controlled way to modify magnetic breakdown and recover nontrivial geometric-phase signatures in kagome systems. Although strain is introduced here primarily as a theoretical tuning parameter to isolate orbit-connectivity effects, the results also suggest a possible route for experimentally testing the proposed mechanism.

\section{\label{sec:level4}Conclusion}
We have shown that the oscillation phase offset is governed not only by the underlying electronic structure, but also by magnetic breakdown between neighboring orbits. Weak orbital hybridization can slightly modify the hybridization gaps at avoided crossings. Small changes in these gaps can substantially modify the fitted Landau-fan phase offset, even though the overall electronic structure remains nearly unchanged. This phase difference is traced to two contrasting magnetic-breakdown regimes. In the strong-breakdown regime, Berry-phase contributions from different orbit segments largely cancel, yielding a near-trivial phase offset. When magnetic breakdown is suppressed, this cancellation is weakened and the nontrivial phase offset becomes visible. Uniaxial strain provides an additional tuning parameter for this process by enlarging the hybridization gaps and reducing inter-orbit tunneling. Although strain is introduced here primarily to isolate orbit-connectivity effects, the results also suggest a possible route for experimentally testing the proposed mechanism. Overall, magnetic breakdown is identified as the key mechanism controlling the measured oscillation phase and provides a plausible explanation for the phase discrepancies reported in kagome metals.

\begin{acknowledgments}
This work was supported by the Central Government Guides Local Science and Technology Development Fund (Grant No.~246Z1027G), the Science Foundation of Hebei Normal University (Grant No.~L2026ZD05), the Science Foundation of Hebei Normal University (Grant No.~L2024J02), and the National Natural Science Foundation of China (Grant No.~11874139).
\end{acknowledgments}

\appendix
\renewcommand{\thefigure}{S\arabic{figure}}
\setcounter{figure}{0}

\section{Momentum-Space Hamiltonian}
\label{appendixA}

For the band-structure and Fermi-surface calculations, we use a symmetrized momentum-space Hamiltonian obtained by absorbing the phase factors $e^{i\mathbf{k}\cdot\mathbf{d}/2}$ into the basis states. In the momentum-space topology analysis, we retain a single projected spin block, whereas the real-space DOS calculation keeps both symmetry-related blocks.

The basis is defined as
\begin{equation}
\Psi_{\mathbf{k}}^\dagger
=
\left(
c_{1\mathbf{k}}^\dagger,\,
c_{2\mathbf{k}}^\dagger,\,
c_{3\mathbf{k}}^\dagger,\,
c_{4\mathbf{k}}^\dagger
\right),
\end{equation}
where the indices $1,2,3$ label the kagome sites and $4$ the central site. The Hamiltonian reads
\begin{equation}
H(\mathbf{k}) = H_{0}(\mathbf{k}) + H_{\text{SOC}}(\mathbf{k}).
\end{equation}

The spin-independent hopping term is
\begin{equation}
\resizebox{0.98\linewidth}{!}{$\displaystyle
H_{0}(\mathbf{k}) = -2
\begin{pmatrix}
0 & t\cos(k_1) & t\cos(k_2) & t_1\cos(k_3) + t_2\cos(k_5) \\
t\cos(k_1) & 0 & t\cos(k_3) & t_1\cos(k_2) + t_2\cos(k_6) \\
t\cos(k_2) & t\cos(k_3) & 0 & t_1\cos(k_1) + t_2\cos(k_4) \\
t_1\cos(k_3) + t_2\cos(k_5) & t_1\cos(k_2) + t_2\cos(k_6) & t_1\cos(k_1) + t_2\cos(k_4) & 0
\end{pmatrix}.
$}
\end{equation}

The projected chiral term is
\begin{equation}
\resizebox{0.98\linewidth}{!}{$\displaystyle
H_{\text{SOC}}(\mathbf{k}) = 2i\lambda
\begin{pmatrix}
0 & \cos(k_2+k_3) & -\cos(k_3-k_1) & 0 \\
-\cos(k_2+k_3) & 0 & \cos(k_1+k_2) & 0 \\
\cos(k_3-k_1) & -\cos(k_1+k_2) & 0 & 0 \\
0 & 0 & 0 & 0
\end{pmatrix}.
$}
\end{equation}

The momentum coordinates are
\begin{align}
k_1 &= k_x, &
k_4 &= \sqrt{3}\,k_y, \nonumber \\
k_2 &= \frac{1}{2}k_x + \frac{\sqrt{3}}{2}k_y, &
k_5 &= -\frac{3}{2}k_x - \frac{\sqrt{3}}{2}k_y, \nonumber \\
k_3 &= -\frac{1}{2}k_x + \frac{\sqrt{3}}{2}k_y, &
k_6 &= \frac{3}{2}k_x - \frac{\sqrt{3}}{2}k_y .
\end{align}

Here $\lambda$ denotes the effective chiral second-neighbor hopping in the projected single-block description.

\section{Block Topology and Orbit-Resolved Phase}
\label{appendixB}

\begin{figure}[htbp]
\centering
\includegraphics[width=0.48\textwidth]{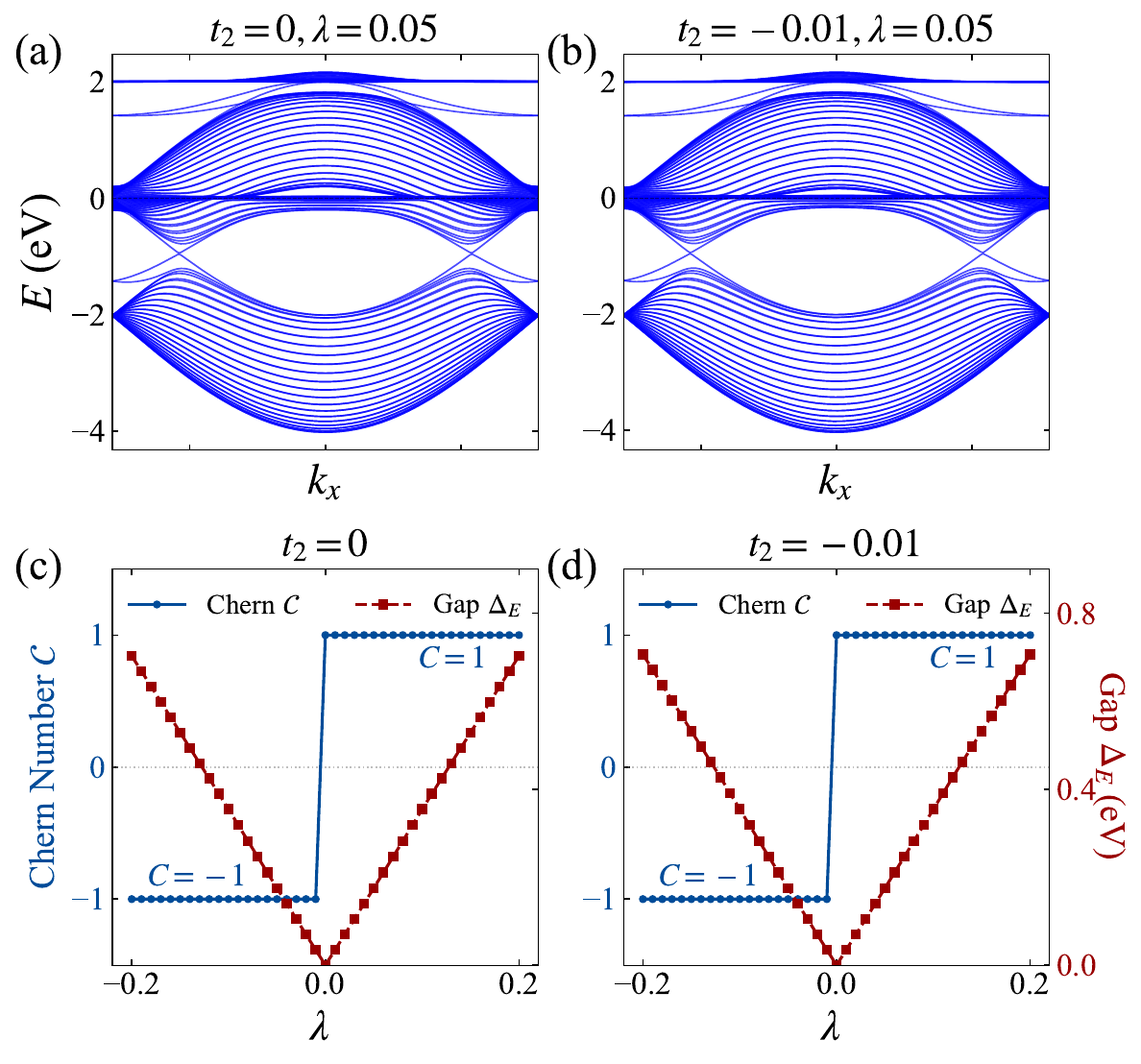}
\caption{Projected single-block band topology for $t_2=0$ and $t_2=-0.01$. (a,b) Nanoribbon spectra at $\lambda=0.05$. (c,d) Block Chern number $\mathcal{C}$ and gap $\Delta_E$ as functions of $\lambda$. The projected block remains topological in both regimes.}
\label{fig:S1}
\end{figure}

This appendix distinguishes the block topology of the projected $4\times4$ Hamiltonian from the orbit-resolved geometric phase entering the quantum-oscillation response. Figure~\ref{fig:S1} shows that the projected block remains topological in both representative regimes $t_2=0$ and $t_2=-0.01$. In the full spinful model, however, the opposite block carries the opposite chirality, so the quoted $\mathcal{C}$ should be understood as a block Chern number rather than the total Chern number.

For quantum oscillations, the relevant quantity is instead the geometric phase accumulated along a specific cyclotron orbit,
\begin{equation}
\phi_{\mathrm{orb}}
=
\oint_{\mathcal C}\mathbf A_m(\mathbf k)\cdot d\mathbf k
=
\iint_{S(\mathcal C)}\Omega_m(\mathbf k)\,d^2k,
\end{equation}
or, in the breakdown regime, the corresponding multi-segment phase of the reconstructed trajectory.
Accordingly, block topology and oscillation phase are related but not equivalent. The extracted phase depends on the Berry curvature sampled by the orbit and on whether magnetic breakdown mixes neighboring bands, allowing partial cancellation. This is why the two hybridization regimes can share the same block topology while exhibiting different phase offsets in the main text.

\section{Recursive Green's Function Scheme}
\label{appendixC}
\begin{figure}[htbp]
    \centering
    \includegraphics[width=0.95\linewidth]{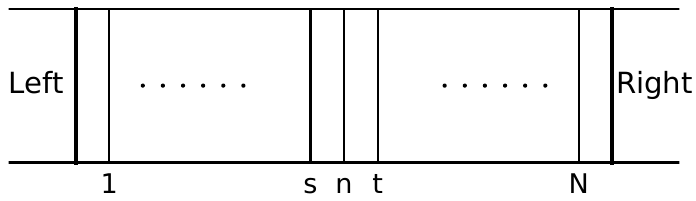}
    \caption{Schematic of the recursive Green's function scheme. The diagonal Green's-function blocks are propagated from the left and right ends toward the central layer $n$, where $G_{nn}^r$ is assembled.}
    \label{fig:S2}
\end{figure}

To compute the density of states at the Fermi energy, we evaluate the diagonal blocks of the retarded Green's function
\begin{equation}
G^r(E) = (E - H + i\eta)^{-1}.
\end{equation}
The lattice is divided into slices and reordered into a block-tridiagonal form, with intralayer block $H_{00}$ and interlayer couplings $H_{01}$ and $H_{01}^\dagger$. The recursive scheme is illustrated in Fig.~\ref{fig:S2}, where the diagonal Green's-function blocks are propagated from the two ends toward the central layer $n$.

Starting from the two outermost slices, we define
\begin{align}
L_1 &= 0, \\
R_N &= 0,
\end{align}
and propagate the left and right self-energies inward as
\begin{align}
L_i &= H_{01}^\dagger \left(E - H_{00} - L_{i-1}\right)^{-1} H_{01}, \qquad i=2,\dots,n, \\
R_i &= H_{01} \left(E - H_{00} - R_{i+1}\right)^{-1} H_{01}^\dagger, \qquad i=N-1,\dots,n.
\end{align}
The diagonal Green's-function block at the central layer is then
\begin{equation}
G_{nn}^{r} =
\left(
E - H_{00} - L_n - R_n
\right)^{-1}.
\end{equation}
Finally, the layer-resolved density of states follows as
\begin{equation}
D_n(E_F) =
-\frac{1}{\pi}\operatorname{Im}\operatorname{Tr}[G_{nn}^{r}(E_F)].
\end{equation}

\section{Supplementary Landau-Fan Data for the Magnetic-Breakdown Branches}
\label{appendixD}
\begin{figure}[htbp]
\centering
\includegraphics[width=0.48\textwidth]{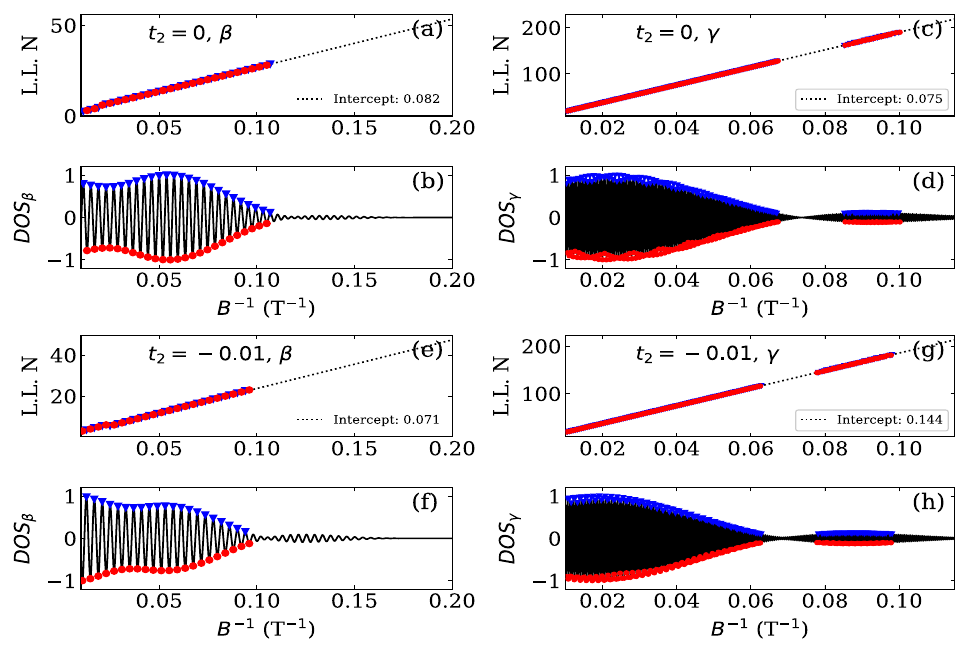}
\caption{Supplementary DOS-based Landau-fan fits for the magnetic-breakdown $\beta$ and $\gamma$ branches at $t_2=0$ and $t_2=-0.01$. All fitted intercepts remain near zero within the same DOS convention used in the main text.}
\label{fig:S3}
\end{figure}

Figure~\ref{fig:S3} shows DOS-based Landau-fan fits for the magnetic-breakdown $\beta$ and $\gamma$ branches in both hybridization regimes. Within the same DOS convention used for the $\alpha$ branch, all fitted intercepts remain close to zero. This supports the main-text conclusion that the pronounced phase evolution is specific to the isolated $\alpha$ orbit rather than to the breakdown trajectories.

\section{Uniaxial Strain Implementation and Supplementary Strain Data}
\label{appendixE}
\begin{figure}[htbp]
\centering
\includegraphics[width=0.48\textwidth]{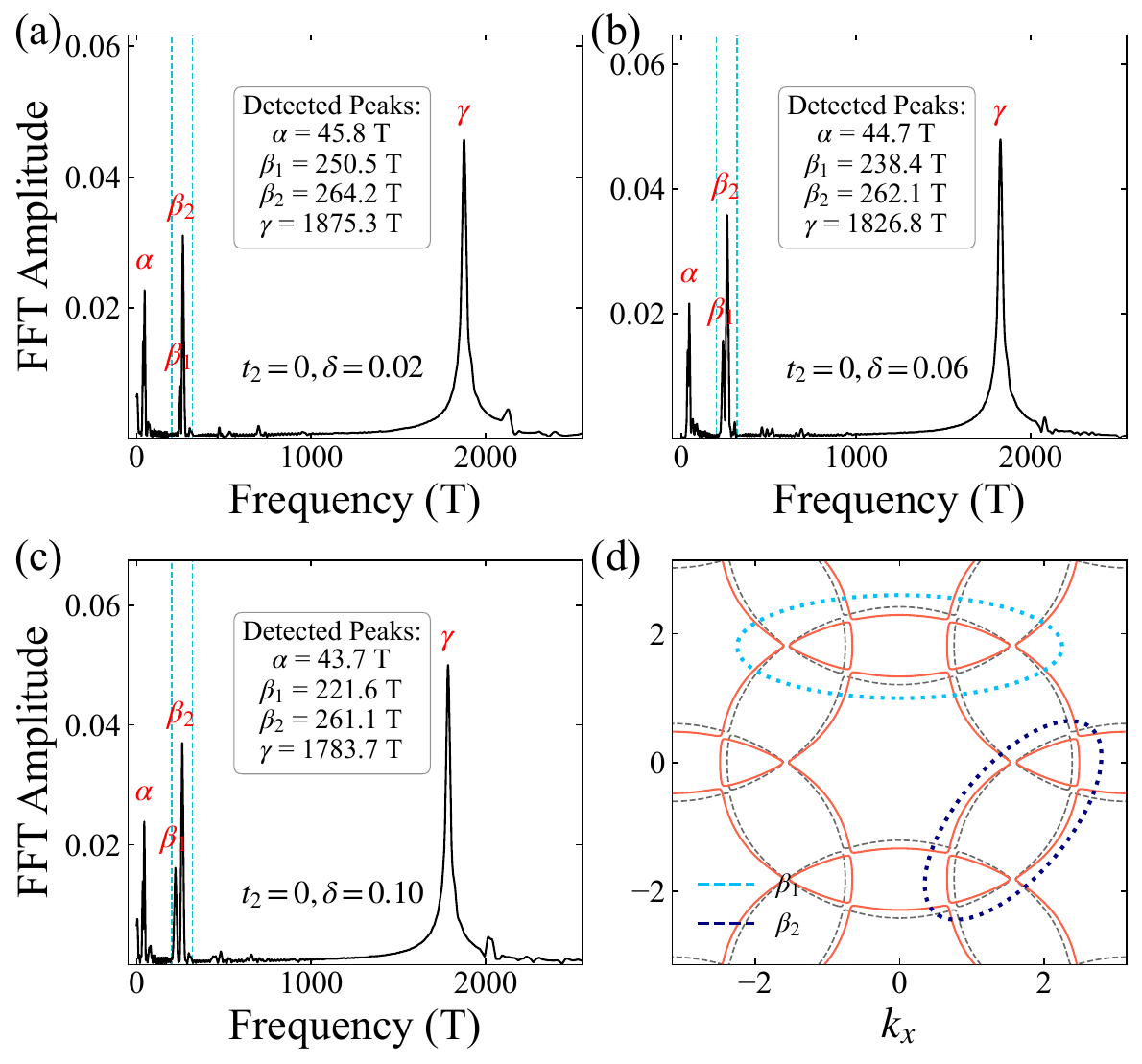}
\caption{Strain evolution and momentum-space origin of the split $\beta$ branches in the strong-breakdown reference regime. Panels (a)-(c) show the splitting of the original $\beta$ peak into $\beta_1$ and $\beta_2$ under strain, with $\beta_1$ shifting downward and $\beta_2$ remaining near $260$~T. Panel (d) shows the corresponding Fermi-surface reconstruction, where the $\beta_1$-related orbit shrinks strongly while the $\beta_2$-related orbit mainly changes shape.}
\label{fig:S4}
\end{figure}

To implement uniaxial strain, we introduce anisotropic hopping renormalizations through the two-dimensional strain tensor
\begin{equation}
\bar{\bar{\varepsilon}} =
\begin{pmatrix}
\varepsilon & 0 \\
0 & -\nu\varepsilon
\end{pmatrix},
\end{equation}
where $\varepsilon$ is the applied longitudinal strain and $\nu$ is the Poisson ratio. For a bond oriented along $\mathbf{n} = (\cos\theta, \sin\theta)$, the relative bond-length variation is
\begin{equation}
\frac{\Delta d}{d_0} = \mathbf{n}^T \bar{\bar{\varepsilon}} \mathbf{n} = \varepsilon (\cos^2\theta - \nu\sin^2\theta).
\end{equation}
Assuming $t = t_0 e^{-\alpha\Delta d/d_0}$ and expanding to linear order, we obtain
\begin{equation}
t(\theta) \approx t_0 \left[ 1 - \delta (\cos^2\theta - \nu\sin^2\theta) \right],
\end{equation}
with $\delta = \alpha\varepsilon$. In the calculations, the representative strained hoppings are
\begin{align}
t_x &= t(1-\delta), \\
t_{\mathrm{diag}} &= t(1-0.25\delta+0.75\nu\delta), \\
t_{2,y} &= t_2(1+\nu\delta).
\end{align}
This anisotropic reconstruction breaks the original $C_6$ symmetry and continuously tunes the magnetic-breakdown network discussed in the main text.

Figure~\ref{fig:S4} provides complementary frequency- and momentum-space evidence for the strain-induced splitting of the original $\beta$ branch. As shown in Fig.~\ref{fig:S4}(a)--(c), strain lifts the near degeneracy and separates the branch into $\beta_1$ and $\beta_2$: $\beta_1$ shifts systematically to lower frequency, whereas $\beta_2$ remains nearly pinned near $260$~T. Figure~\ref{fig:S4}(d) shows the corresponding Fermi-surface reconstruction, where the $\beta_1$-related orbit shrinks substantially while the $\beta_2$-related orbit mainly changes shape with only a weak change in enclosed area. Together, these results support the interpretation that the intermediate-strain regime is governed by orbit-selective splitting and the resulting beating under partial magnetic breakdown.

\section*{Data Availability}
The numerical data and computation codes that support the findings of this study are available from the corresponding authors upon reasonable request.

\end{document}